\documentclass[aps,prb,preprint,showkeys]{revtex4-1}
\usepackage{graphicx}

\usepackage{amsmath,amssymb,amsfonts}
\usepackage{xcolor}%
\usepackage[breaklinks,colorlinks = true,linkcolor = blue,urlcolor=blue,citecolor=blue]{hyperref}

\begin{document}

\title{Graph Neural Network Approach to Predicting Magnetization in Quasi-One-Dimensional Ising Systems}

\author{V. Slavin}
\affiliation{B. Verkin Institute for Low Temperature Physics and Engineering of the National Academy of Sciences of Ukraine, Nauky Ave., 47, Kharkiv, 61103, Ukraine\\
email: slavin@ilt.kharkov.ua}

\author{O. Kryvchikov}
\affiliation{B. Verkin Institute for Low Temperature Physics and Engineering of the National Academy of Sciences of Ukraine, Nauky Ave., 47, Kharkiv, 61103, Ukraine\\
email: kryvchikov@ilt.kharkov.ua}

\author{D. Laptev}
\affiliation{B. Verkin Institute for Low Temperature Physics and Engineering of the National Academy of Sciences of Ukraine, Nauky Ave., 47, Kharkiv, 61103, Ukraine\\
email: laptev@ilt.kharkov.ua}

\begin{abstract}
We present a graph-based deep learning framework for predicting the magnetic properties of quasi-one-dimensional Ising spin systems. The lattice geometry is encoded as a graph and processed by a graph neural network (GNN) followed by fully connected layers. The model is trained on Monte Carlo simulation data and accurately reproduces key features of the magnetization curve, including plateaus, critical transition points, and the effects of geometric frustration. It captures both local motifs and global symmetries, demonstrating that GNNs can infer magnetic behavior directly from structural connectivity. The proposed approach enables efficient prediction of magnetization without the need for additional Monte Carlo simulations.
\end{abstract}

\keywords{Ising Model, Monte-Carlo Method, Graph Neural Network, Deep Neural Network.}

\maketitle

\section{Introduction}\label{sec1}

A graph is a data structure defined by a set of vertices $\mathcal{V}$ and a set of edges $\mathcal{E}$, denoted as $G = (\mathcal{V}, \mathcal{E})$. In recent years, graph-based methods have been extensively applied to molecular systems, as molecules can be naturally represented as graphs. This has led to the rapid development of Graph Neural Networks (GNNs), a subclass of deep neural networks (DNNs) designed to operate on graph-structured data. GNNs have demonstrated strong performance in various applications, including the prediction of physicochemical properties such as hydrophobicity~\cite{shang2018edge,wang2019spatial,becigneul2020optimal} and toxicity~\cite{xu2017toxicity,withnall2020edge,yuan2020structpool,hu2019pretraining}.

In this work, we extend this approach to problems in statistical physics, specifically to a class of one-dimensional Ising models. The Hamiltonian of the system is given by:

\begin{equation}
\mathcal{H} = J\sum\limits_{\vec{r_i}, \vec{R}} S_{\vec{r_i}} S_{\vec{r_i}+\vec{R}} + h\sum\limits_{\vec{r_i}}  S_{\vec{r_i}}, \label{eq:original_H}
\end{equation}

Here, $J$ denotes the coupling between Ising spins $S$, and $h$ is the external magnetic field. We consider the antiferromagnetic case with $J = 1$. The interactions are short-ranged, satisfying $\|\vec{R}\| \ll$ system size. The underlying lattice geometry is arbitrary but assumed to be periodic. The spin position vector is given by $\vec{r}_i = 
\begin{pmatrix}
x \\
y 
\end{pmatrix}$, where the transverse component satisfies $y \lesssim R$. The system is thus quasi-one-dimensional: it extends along x axis, while the transverse width remains finite.

The Hamiltonian of the Ising model can be naturally expressed as a sum over the vertices of a graph, where the geometry of the lattice unit cell determines the edge structure. Due to this formulation, the system admits equivalent representations via graph isomorphism: graphs with identical connectivity but differing layouts are indistinguishable. This allows the Hamiltonian to be mapped onto a one-dimensional form, facilitating further analysis.

\begin{figure}[ht]
\centering
\includegraphics[width=0.8\textwidth]{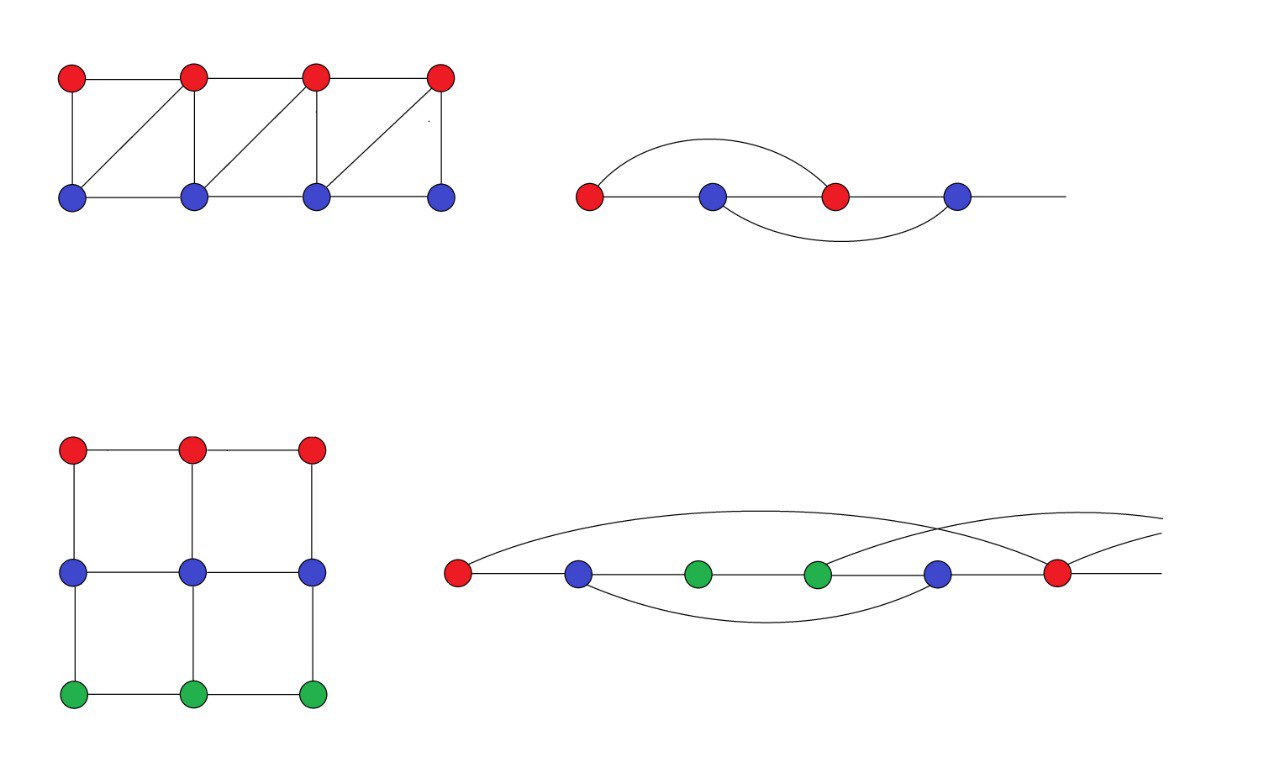} 
\caption{Isomorphic representations of two interaction graphs. The top shows a nearest-neighbor interaction chain, while the bottom illustrates a three-leg rectangular ladder. }
\label{fig:graph}
\end{figure}

The mapping is achieved by partitioning the Hamiltonian into $N$ smaller blocks that repeat every $M$  sites. The total number of spins is then given by $NM$. Each spin located at position $\vec{r_i}$ is then relabeled as $S_{\vec{r_i}} \rightarrow S_{j + iM}$, where $j = 0, \dots, M - 1$ is the intra-block index and $i = 0, \dots, N - 1$ is the block index.

\begin{equation}
\mathcal{H} =\sum\limits_{i=0}^{N}\left(\sum\limits_{j\ne k}^{M}{W}_{j,k}{S}_{j+iM}{S}_{k+iM} + \sum\limits_{j\ne k}^{M}\tilde{W}_{j,k}{S}_{j+iM}{S}_{k+(i+1)M} + \sum\limits_{j}^{M}h{S}_{j + iM}\right).\label{eq1}
\end{equation}

The Hamiltonian~\eqref{eq1} is fully equivalent to the original form~\eqref{eq:original_H}, but mapped onto a one-dimensional representation. The total energy is expressed as a sum over $N$ identical structural units, each containing $M$ spins. Within each unit, the spins $S_j$ and $S_k$ interact through the coupling matrix $W_{j,k}$, which encodes intra-block interactions. The matrix $\tilde{W}_{j,k}$ accounts for couplings between spins located in adjacent blocks.

Both $W_{j,k}$ and $\tilde{W}_{j,k}$ are symmetric, as the spin interaction is invariant under the exchange of indices. In this study, we restrict the matrix elements of $W_{j,k}$ and $\tilde{W}_{j,k}$ to values $0$ or $J$, corresponding to the absence or presence of antiferromagnetic interaction between the respective spins. This formalism generalizes the classical Ising chain by permitting arbitrary coupling patterns within and between units, while still disallowing long-range interactions that span distant blocks.

To obtain magnetization curves for a given configuration of $W_{j,k}$ and $\tilde{W}_{j,k}$, we employ the classical Monte Carlo simulations.

The Hamiltonian~\eqref{eq1} can be naturally interpreted as an undirected graph, where each vertex corresponds to a spin, and edges represent pairwise interactions. Once simulation data are collected, we use the resulting magnetization curves to train a neural network. The network learns the mapping from the interaction graph, defined by $W_{j,k}$ and $\tilde{W}_{j,k}$, to the corresponding magnetization behavior. After training, the model can predict magnetization curves for previously unseen graph configurations without requiring additional Monte Carlo simulations.

\section{The model}\label{sec2}

To directly predict the magnetization from the structure of the interaction graph—without relying on computationally expensive Monte Carlo simulations—we employ Graph Neural Networks (GNNs).  Modern Graph Neural Networks (GNNs) are well-suited for modeling spin systems such as the Ising model, where the underlying structure of spin interactions can be naturally represented as a graph. 

Let $G = (\mathcal{V}, \mathcal{E})$ denote the input graph, where and each edge $ e_{ij} \in \mathcal{E} $ encodes the interaction (i.e., non-zero coupling \( W_{ij} \) or \( \widetilde{W}_{ij} \)) between spins $ S_i $ and $ S_j $. Each node $u_i \in \mathcal{V}$ in the interaction graph may carry various types of information, such as structural or topological features. Since the actual spin states are unknown, the $f_{\theta}$ is a mapping from the graph topology and encoded edge features to macroscopic observables like magnetization. In other words, we want to obtain a relation like so:

\begin{equation}
M(h) = f_{\theta}(G)\end{equation}

where $G = (\mathcal{V}, \mathcal{E})$ is the input graph, and $f_{\theta}$ is the GNN model parameterized by learnable weights $\theta$

Each node feature vector is initialized as
\begin{equation}
u_i^{(0)} = \begin{pmatrix} i \\ k_i \end{pmatrix},
\end{equation}
where $i$ denotes the index of the spin within the unit cell (e.g., 1, 2, 3, etc.), and $k_i$ represents the number of neighboring spins connected to node $u_i$. The edge features \( a_{ij} \) are derived from the interaction strength (e.g., \( W_{ij} = 0 \) or \( 1 \)).

The GNN operates through a series of message-passing layers. At each layer \( k \), node \( u_i \) aggregates information from its neighbors via a message function:

\begin{equation}\label{eq:message}
m_i^{(k)} = \sum_{i \neq j} \phi_m^{(k)}\left(u_i^{(k-1)}, u_j^{(k-1)}, e_{ij}\right),
\end{equation}

where the summation goes over the neighbors of \( u_i \), and \( \phi_m^{(k)} \) is a learnable function that determines how messages are computed. The aggregation in~\eqref{eq:message} employs uniform summation over neighbors without attention, consistent with the Graph Convolutional Network (GCN) framework \cite{Kipf2016}. 

The aggregated message is then used to update the node’s feature:

\begin{equation}\label{eq:update}
u_i^{(k)} = \phi_u^{(k)}\left(u_i^{(k-1)}, m_i^{(k)}\right),
\end{equation}

where $\phi_u^{(k)}$  is a learnable update function. This process is repeated for $K$ layers, allowing spin states to be influenced by increasingly distant neighbors in the interaction graph. Steps~\eqref{eq:message}–\eqref{eq:update} are crucial for enabling each node to aggregate information from its local neighborhood and capture structural properties of each graph. 

After  $k$ message-passing layers, the feature vector  $u_i^{(k)}$  encodes information about its neighborhood up to $k$ hops. The number of layers  $K$  in the GNN architecture is a hyperparameter. Increasing $K$ allows the network to capture long-range dependencies in the graph. However, choosing a too-large value of $K$ may lead to the oversmoothing problem~\cite{li2018deeper,oono2020graph}, in which node embeddings across the graph become nearly indistinguishable.

 In our case, the number of nodes varies from graph to graph, since each graph represents a different configuration of a magnetic system with a different number of spins. However, the model must produce a fixed-size output (e.g., a scalar magnetization value) regardless of the graph's size. To predict properties of the entire system, a readout function \( R(\cdot) \) is applied after the final layer to produce a graph-level representation:

\begin{equation}\label{eq:readout}
u_G = R\left(u_i^{(K)}\right).
\end{equation}

$u_G$ is a fixed-size vector known as the graph embedding, which encodes aggregated information from all nodes $\mathcal{V}$. A graph embedding is a continuous vector representation of a graph that captures its structural and attribute information. These embeddings reside in a latent space, a lower-dimensional representation space where graph similarity corresponds to geometric proximity.
Using~\eqref{eq:readout}, a scalar parameter can be defined that describes the similarity between two graphs. This parameter is an angle between two vectors:
\begin{equation}
S =  \frac{\left(u_{G,1}, u_{G,2}\right)}{\|{u_{G,1}}\|\|{u_{G,2}}\|}
\end{equation}
This measure reflects the alignment of two graph-level embeddings ${u}_{G_1}, {u}_{G_2}$. 

Several studies have proposed methods for learning and comparing graph representations via embedding-based approaches.  In~\cite{bai2019simgnn}, a neural model with attention mechanisms is used to identify important nodes and estimate similarity between graphs. The model of ~\cite{li2019graphsim} direct node-level set matching using multiscale convolution, is performed bypassing the need for fixed-size graph vectors. In the study ~\cite{zhang2021graph}, a symmetric attention-based similarity metric learning approach is applied to compare graph embeddings.

To obtain a fixed-length vector representation that preserves permutation invariance and captures the distributional properties of the node embeddings ${u}_i$, we employ the Set2Set mechanism~\cite{vinyals2015order}. This method maps unordered sets of arbitrary size into fixed-dimensional embeddings, reflecting key statistical and structural characteristics such as clustering, spread, and overall geometry. As a result, structurally similar graphs are projected to nearby points in the embedding space, even when differing in size or topology, while structurally distinct graphs produce more divergent outputs.

We now provide a more detailed description of how this mechanism operates. The Set2Set readout is computed through an iterative attention process controlled by an LSTM over $T$ steps. The Long Short-Term Memory (LSTM) network~\cite{hochreiter1997long,gers2000learning,greff2016lstm} enables the model to capture long-range dependencies and aggregate global information from the set of node features, producing a permutation-invariant vector ${u}_G$ that summarizes the entire graph.

At each iteration $ t$ , the following operations are performed:
LSTM update:
\begin{equation}
    ({q}_t, {c}_t) = \mathrm{LSTM}({r}_{t-1}, ({q}_{t-1}, {c}_{t-1})), {q}_0 = {0}, \quad {c}_0 = {0}, \quad {r}_0 = {0} 
\end{equation}

The vector ${q}_t$ represents the hidden state of the controller and determines the current attention focus or importance of different parts of the graph. The memory cell state $c_t$ retains long-term contextual information across iterations.

Normalization of weights:
\begin{equation}
    \alpha_i^{(t)} = \frac{\exp\left( {u}_i^\top {q}_t \right)}{\sum\limits_{j=1}^n \exp\left( {u}_j^\top {q}_t \right)}, \quad \text{for } i = 1, \dots, n
\end{equation}

At each iteration step, an attention weight vector is computed to assign importance scores to different parts of the graph. 
\begin{equation}
    {r}_t = \sum_{i=1}^n \alpha_i^{(t)} {u}_i
\end{equation}

The node embeddings are then aggregated using these weights to produce a context vector that summarizes the most relevant structural information. After \( T \) iterations, the final graph embedding is obtained:
\begin{equation}
{u}_G = [{r}_1 \| {r}_2 \| \cdots \| {r}_T] 
\end{equation}

The final stage of the model is a multi-layer perceptron (MLP), a simple subclass of deep neural networks (DNNs) composed of multiple fully connected layers. Each layer consists of neurons, and all neurons in one layer are connected to every neuron in the subsequent layer. Given a graph-level embedding ${u}_G$, the MLP with $L$ layers transforms it as follows:

\begin{equation}
{m}^{(0)} = f{u}_G 
\end{equation}
\begin{equation}
{m}^{(l)} = \sigma \left( {W_{\text{MLP}}}^{(l)} {h}^{(l-1)} + {b_{\text{MLP}}}^{(l)} \right), \quad \text{for } l = 1, \dots, L-1 
\end{equation}
\begin{equation}
\hat{{y}} = {W_{\text{MLP}}}^{(N)} {m}^{(N-1)} + {b_{\text{MLP}}}^{(N)}
\end{equation}

where ${W_{MLP}}^{(l)}$  and ${b_{MLP}}^{(l)}$ are the weight matrix and bias vector of layer \( l \), respectively, and $\sigma(\cdot)$ is a non-linear activation function.

Let ${h} = (h_1, h_2, \dots, h_k)$ be a set of discrete external field values. The output vector ${M}({h}) $ represents the predicted magnetization of the Hamiltonain ~\eqref{eq1} corresponding to each value $h_i$. 

\begin{equation}
M(h) =
\begin{pmatrix}
M(h_1) \\
M(h_2) \\
\vdots \\
M(h_k)
\end{pmatrix}
\end{equation}

Each component $\hat{y}_i$  of the output vector corresponds to the predicted value for the discrete field $h_i$, i.e.,

\begin{equation}\label{eq:M_h}
M(h_i) = \hat{y}_i, \quad \text{for } i = 1, \dots, k.
\end{equation}

Thus, we can construct the following computational pipeline from the input graph to the output:

\[
\mathcal{V, E}\longrightarrow\text{GCN} \longrightarrow \text{Set2Set} \longrightarrow \text{MLP} \longrightarrow \hat{y_i}
\]

The output vectors $\hat{y}_i$ are used to define the loss function $\mathcal{L}$, which quantifies the discrepancy between predicted and reference magnetization curves. The model is trained by minimizing this loss, with performance improving as 
$\mathcal{L}$ decreases. Specifically, the mean squared error (MSE) is employed as the loss function:

\begin{equation}
\mathcal{L} = \frac{1}{k} \sum_{i=1}^k \left( \hat{y}_i - \tilde{M}_i \right)^2
\end{equation}
where $\tilde{M}_i$ denotes the magnetization values computed from Monte Carlo simulations, and $\hat{y}_i$ are the model predictions corresponding to external field values $h_i$.

Training\cite{Nielsen2015} refers to the iterative process of adjusting the model parameters to minimize a predefined loss function, which quantifies the discrepancy between the predicted output $\hat{y}_i$ and the true target $\tilde{M}_i$. During training, the model optimizes its parameters $\theta$ to minimize $\mathcal{L}$.

\begin{equation}
\theta = \left\{ {W}_{\text{MLP}}, {b_{\text{MLP}}}, {W}_{\text{Set2Set}}, {b_{\text{Set2Set}}}, {W}_{\text{GCN}}, {b_{\text{GCN}}} \right\}
\end{equation}

At each training step, the gradient of the loss function with respect to the model parameters \( \theta \) is computed:

\begin{equation}
\nabla_\theta \mathcal{L} = \frac{\partial \mathcal{L}}{\partial \hat{{y}_i}} \cdot \frac{\partial \hat{{y_i}}}{\partial \theta}
\end{equation}

In gradient descent, the model parameters are updated by stepping in the direction opposite to the loss gradient. Concretely, if $\theta_t$ denotes the parameters at iteration $t$ and $\mathcal{L}(\theta_t)$ is the loss, the basic update is
\[
\theta_{t+1} = \theta_t - \eta \, \nabla_\theta \mathcal{L}(\theta_t),
\]
where $\eta $ is the learning rate. The structure and evolution of the weight matrices $W_{\text{MLP}}$, $W_{\text{Set2Set}}$, $W_{\text{GCN}}$ during training play a crucial role in the dynamics and generalization ability of the network. Recent research \cite{Pastur2023} has shown that the spectral properties of these matrices, particularly the distribution of their singular values, can strongly influence the training process. 

Backpropagation \cite{Rumelhart1986,Goodfellow2016} is the algorithm used to compute gradients of the loss function with respect to the model parameters. It applies the chain rule of calculus to efficiently propagate error signals from the output layer backward through the network. This allows the model to determine how each parameter contributed to the error and update it accordingly.
This involves applying the chain rule to propagate gradients from the output layer back through the MLP, then through the Set2Set mechanism, and finally to the GCN layers:

\[
\text{GCN} \longleftarrow \text{Set2Set} \longleftarrow \text{MLP}
\]

When training is completed, the model can predict the  magnetization curve  $M(h)$ for any input graph. The quality of these predictions strongly depends on the representativeness and diversity of the training dataset, as well as on the efficiency of the training procedure, including hyperparameter tuning and optimization strategies.

\section{Trainig and tuning}
\label{sec3}

The dataset was obtained by collecting the dependence of magnetization on external magnetic field for 80 distinct samples. Each sample was based on a structural unit containing between 1 and 16 spins, with interactions defined by a specific input graph intra-unit and unit-to-unit connectivity pattern. The unit was translated 250 times to form the entire lattice.  On each lattice, Monte-Carlo calculations using the Metropolis algorithm were performed. The inverse temperature was $\beta = 5.0$. The field interval $-4<h<4$ was used. The Monte-Carlo calculations for each field were performed in 12 loops. Each loop consisted of $10^6$ steps. The final magnetization was computed as the average over loops. As a result, for each of the 80 samples, magnetization was obtained as a function of the external field and the corresponding input graph. Together, these formed the dataset used for training and evaluation.

In our problem, the target output is a magnetization curve $M(h)$, which is known to be an odd, non-decreasing function of the external field $h$. That is, it satisfies:

\begin{equation}
-M(-h) = M(h), \qquad \frac{dM}{d|h|} \geq 0.
\end{equation}

However, the symmetry properties of the model output function are not explicitly encoded or enforced in the model architecture.
One naive approach is to augment the loss function $\mathcal{L}$ with penalty terms that increase the loss when $\hat{M}(H)$ is not monotonic or not symmetric:

\begin{equation}
\mathcal{L}_{\text{total}} = \mathcal{L}_{\text{MSE}} + \lambda_1 \cdot \mathcal{L}_{\text{asym}} + \lambda_2 \cdot \mathcal{L}_{\text{nonmono}},
\end{equation}

where $\mathcal{L}_{\text{asym}}$ penalizes $\hat{M}(H) \neq -\hat{M}(-H)$, and $\mathcal{L}_{\text{nonmono}}$ penalizes  regions where $\hat{M}(H)$ decreases. However, it was empirically observed that such penalties degrade overall performance and lead to a loss of accuracy in the regression task.

Let’s explore an alternative approach using cumulative growth with plateau control. Let $z_i = \hat y_i$ be the output of the MLP block for all $h_i>0$. Let's define a positive growth vector $\Delta^+$ as follows:

\begin{equation}
\Delta^+_i = \log\left(1 + e^{z_i}\right).
\end{equation}

 $\Delta^+_i > 0$ for all $z_i$.
 
The goal is to construct an output vector $\hat y_{\text{right}} \in [0,1]^k$ that depends on $h$ monotonically, and has interpretable growth steps.  
\begin{equation}
\hat y_{\text{right}} = \frac{\left[ f(\Delta_1), f(\Delta_1) + f(\Delta_2), \ldots, \sum_{i=1}^{k} f(\Delta_i) \right]}{\sum_{i=1}^{k} f(\Delta_i)},
\end{equation}

Lets define the  vector $f(\Delta_i)$ as follows:

\begin{equation}
f(\Delta_i)= \Delta^+_i  \delta_{[\sum\Delta^+_i> \tau]}
\end{equation}

 The Kronecker symbol $\delta_{[\sum\Delta^+_i > \tau]}$ introduces a  plateau mask to control step-wise behavior of $y_{\text{right}}(h)$:

\begin{equation}
 \delta_{[z_i > \tau]} = 
\begin{cases}
1, & \text{if } \sum\Delta^+_i > \tau, \\
0, & \text{otherwise}.
\end{cases}
\end{equation}

where $\tau$ is a hyperparameter of the plateau threshold, and the summation $\sum\Delta^+_i$ goes on the current plateau.

To ensure symmetry, we define the left-side decrease as:

\begin{equation}
\hat y_{\text{left}} = -\text{flip}(y_{\text{right}}),
\end{equation}

The final smoothing function would be like this:

\begin{equation}
\hat{y}_{\text{smooth}} = \left[ y_{\text{left}}, 0, y_{\text{right}} \right] .
\end{equation}

This ensures that the resulting function is odd and monotonic.

The final predicted output is a weighted sum:

\begin{equation}
\hat{y} = (1 - \alpha) \cdot \hat{y}_i + \alpha \cdot \hat{y}_{\text{smooth}},
\end{equation}

allowing the model to combine learned sharp responses with plateau-aware cumulative behavior. Even a small $\alpha \approx 0.05$ yields smoother results and a more relevant $M(h)$ curve.

In machine learning, hyperparameters as $\alpha$ are external configuration values that are not learned from the training data but are instead set before the training process begins. They define the structure of the model and influence the training dynamics. Selecting appropriate hyperparameters is crucial for achieving good performance and model generalization.

For the GCN $\rightarrow$ Set2Set $\rightarrow$ MLP architecture \cite{Wang2022gnnmolecules}, the key hyperparameters include: learning rate, number of layers in the MLP, number of hidden units per MLP layer, number of GCN layers, choice of activation function, batch size and number of training epochs, plateau threshold, blending coefficient.
Finding the optimal combination of hyperparameters is a non-trivial task. One commonly used method is grid search \cite{Zahedi2021Search}, where a predefined set of possible values is exhaustively explored for each hyperparameter. The model is trained for each combination, and the one yielding the best validation performance is selected. However, grid search becomes inefficient as hyperparameters grow, especially for composite models like GCN+Set2Set+MLP. In the case of composite models, manual hyperparameter tuning may be more effective, as it enables sequential adjustments with immediate feedback on performance. Furthermore, while the loss function measures the discrepancy between the predicted and target outputs, it typically does not reflect the physical plausibility of the predictions.

\section{Results}
\label{sec4}
 
A separate testing dataset was constructed using the same procedure as the training dataset. Each entry comprised an input interaction graph and the corresponding magnetization curve $M(h)$. Importantly, the graphs included in the testing set were entirely disjoint from those used during training, ensuring that the model had no prior exposure to their structural features and allowing for an unbiased evaluation of generalization performance.

To quantitatively assess the agreement between predicted and reference magnetization curves obtained via Monte Carlo simulations, the following composite error metric was employed:
 
 \begin{equation}
\text{E} = \frac{1}{2} \cdot \frac{1}{k} \sum_{i=1}^{k} \left| M^\mathrm{MC}_i - M^\mathrm{pred}_i \right| + 
\frac{1}{2} \cdot \frac{1}{k} \sum_{i=1}^{k} \left| \frac{dM^\mathrm{MC}}{dh_i} - \frac{dM^\mathrm{pred}}{dh_i} \right|
\end{equation}
The comparison is performed over \( k \) discrete external field values, as defined in~\eqref{eq:M_h}. This metric jointly quantifies the absolute deviation in magnetization values and discrepancies in the local slope, thereby capturing both vertical and structural errors in the predicted curves.

When evaluated on previously unseen test graphs, the model demonstrated qualitatively accurate and, in many cases, quantitatively precise agreement with the Monte Carlo results. The minimum observed error was \( E = 0.045 \). Efforts to correlate prediction accuracy with structural properties of the graphs such as through clustering and related analytical techniques did not yield any consistent patterns. Thus, we cannot reliably predict which graph configurations have better performance on our model. The largest error, \( E = 0.389 \), was observed for the two-leg rectangular lattice. This comparatively poor performance may be attributed to the structural regularity and low complexity of this geometry, which could limit the GNN’s capacity to extract informative interaction features. Alternatively, it may arise from imbalances or sparsity in the training dataset.

To investigate the structural organization of the dataset, we employed the t-distributed Stochastic Neighbor Embedding (t-SNE) algorithm, a non-linear dimensionality reduction technique that projects high-dimensional data into a low-dimensional space while preserving local similarity relationships~\cite{vanDerMaaten2008}. Applied to the graph representations, t-SNE allows for qualitative assessment of clustering tendencies and structural similarities between graph instances.

\begin{figure}[ht]
  \centering
  \includegraphics[width=0.8\textwidth]{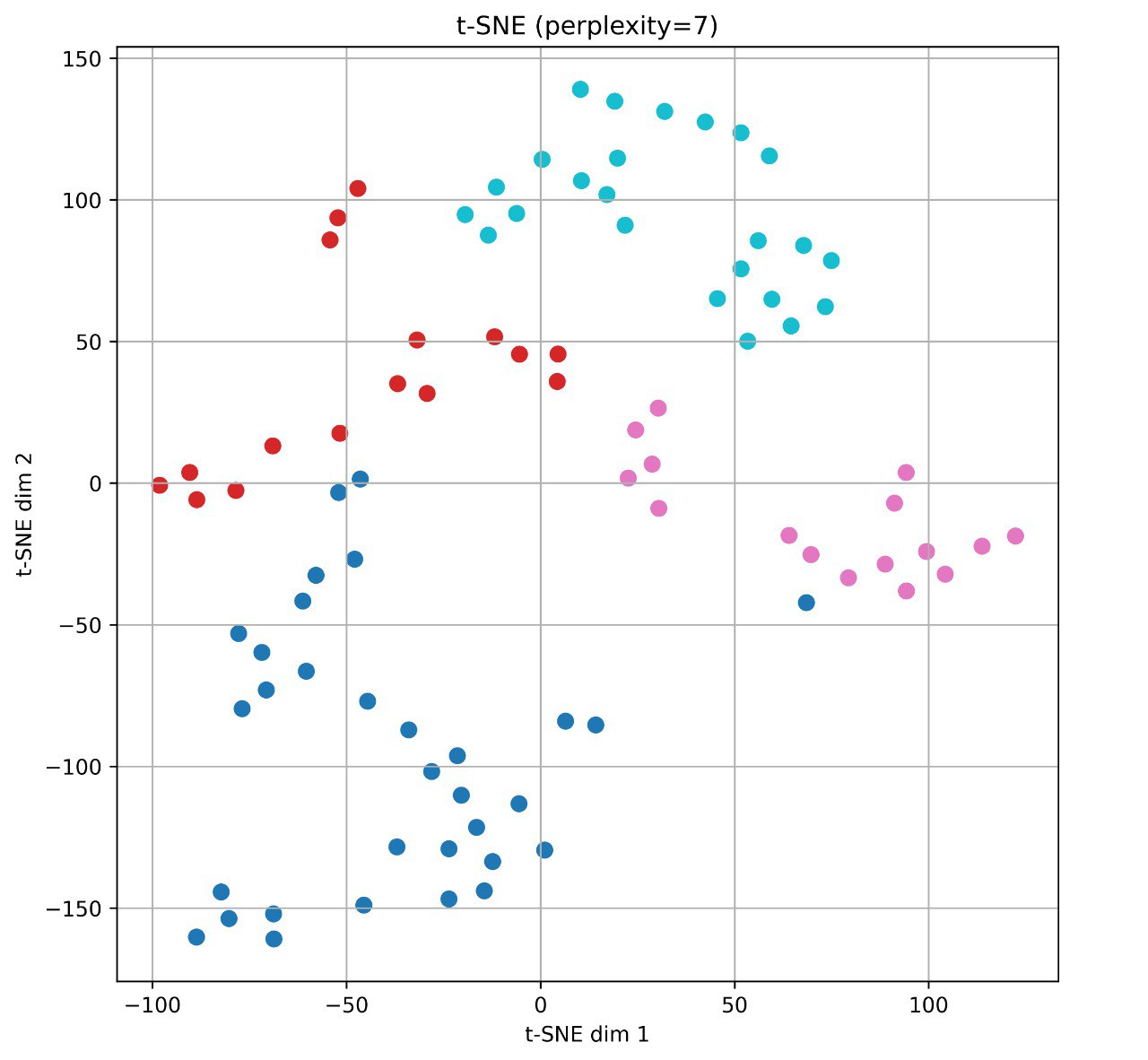} 
  \caption{t-SNE projection of the dataset illustrating four distinct clusters. The light blue region corresponds to rectangular-like lattices, the pink region to chain-like non-frustrated lattices, and the red and blue regions represent different types of frustrated lattices.}
  \label{fig:tsne}
\end{figure}

As shown in Fig.~\ref{fig:tsne}, the two-dimensional t-SNE embedding reveals four loosely separated regions, each corresponding broadly to particular classes of graph structures. The light blue region predominantly contains rectangular-like lattices, which exhibit a high degree of regularity and symmetry. The pink region corresponds mostly to chain-like, non-frustrated lattices. The red and dark blue regions mainly consist of frustrated lattices with more complex connectivity, including motifs such as odd cycles and side couplings.

However, it should be noted that frustration may still be present to some extent within each of these regions, and the exact criteria by which t-SNE clusters the data are not fully determined. The observed clustering reflects patterns in the high-dimensional input space but does not necessarily correspond to a unique physical classification. Nevertheless, this visualization suggests that the learned graph embeddings capture structural distinctions that may be relevant for predicting magnetic behavior.

Despite these limitations, the model was able to consistently capture essential qualitative characteristics of the magnetization curves across most test cases, including the number and height of magnetization plateaus, as well as the positions of transition points.

\begin{figure}[ht]
  \centering
  \includegraphics[width=0.8\textwidth]{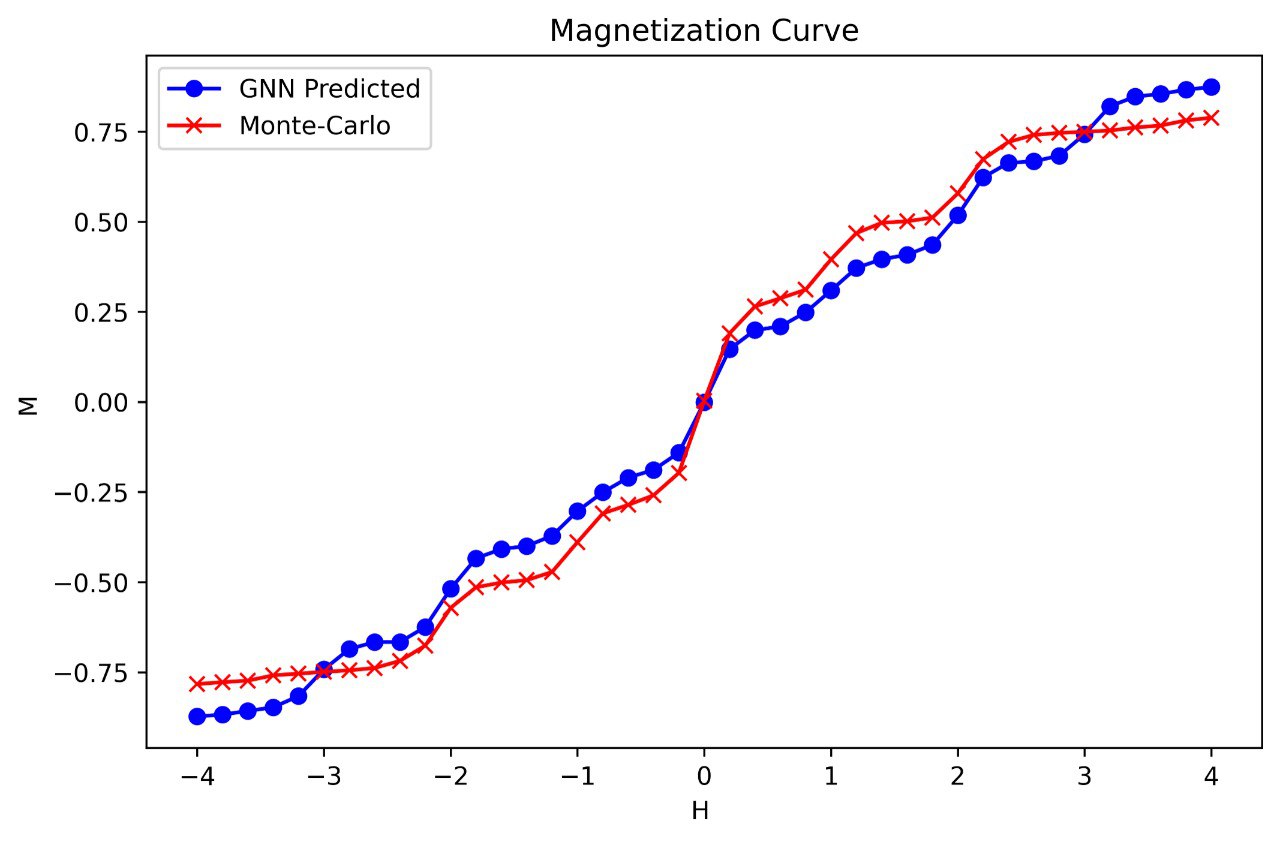} 
  \caption{The dependence of the magnetization on the external field. Tested Monte Carlo (red) and predicted (blue). $E$ = 0.17. }
  \label{fig:magnetization}
\end{figure}
\begin{figure}[htbp]
  \centering
  \includegraphics[width=0.8\textwidth]{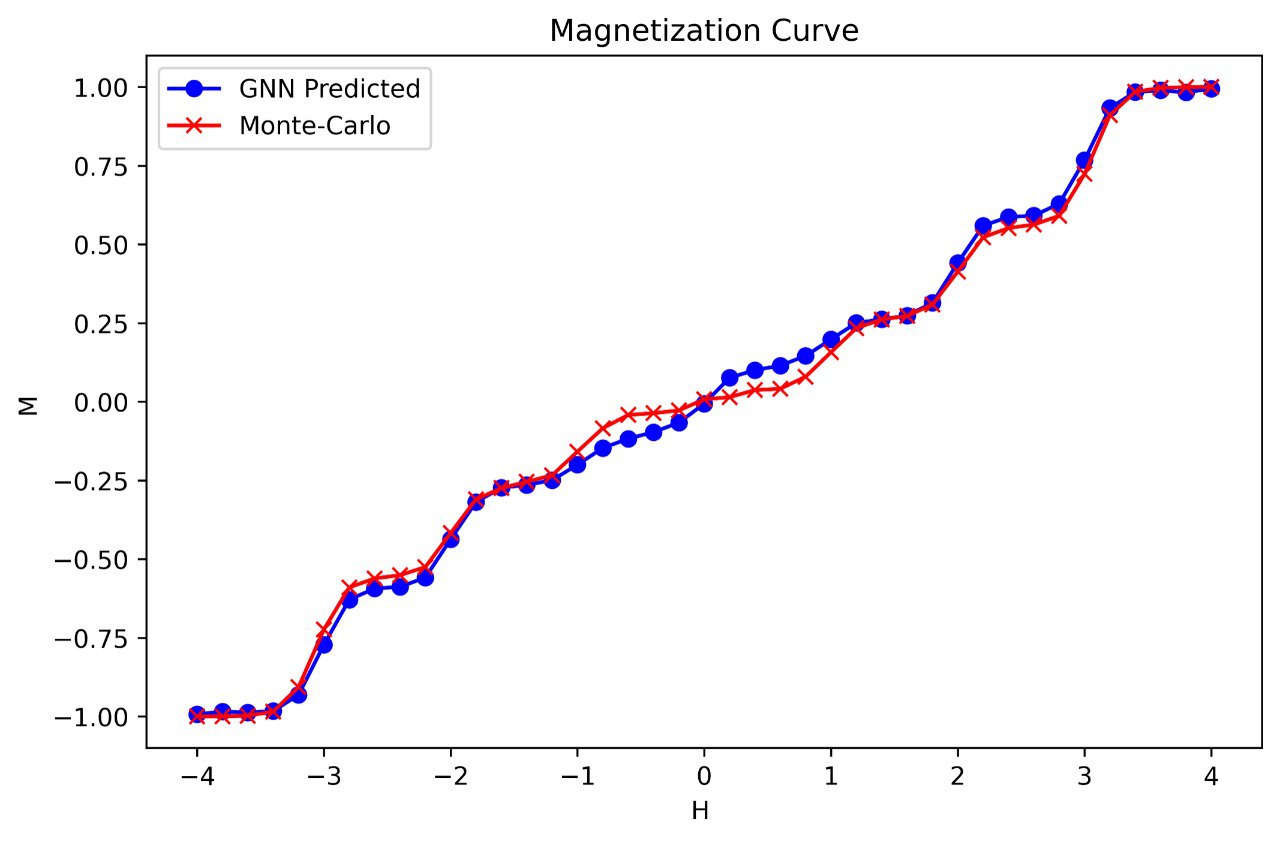} 
  \caption{The dependence of the magnetization on the external field. Tested Monte Carlo (red) and predicted (blue). $E$ = 0.08.}
  \label{fig:magnetization2}
\end{figure}

Structural information was systematically reflected in the predictions produced by the model. Graphs characterized by higher degrees of symmetry typically yielded smoother magnetization profiles, whereas those with sparse or irregular connectivity exhibited more abrupt transitions.

These trends are in agreement with established theoretical results for one-dimensional spin systems. Prior studies have shown that non-frustrated configurations tend to display well-defined magnetization plateaus and sharp field-induced transitions~\cite{Zvyagin2024,Zvyagin2025,Zvyagin2023}. In contrast, frustrated geometries—especially those involving antiferromagnetic couplings on odd-length cycles or exhibiting compositional disorder—give rise to more intricate magnetization behavior, including plateau flattening, delayed saturation, and suppression of total magnetization~\cite{Laptiev2024,Slavin2021,Kryvchikov2022,Cher:2019,Slav:2003}. Remarkably, the proposed model successfully captures these qualitative distinctions based solely on structural input, underscoring its capacity to extract and internalize essential physical features governing low-dimensional magnetic phenomena.

Furthermore, the network demonstrates sensitivity to local structural motifs that are known to introduce frustration, such as side couplings and cycles of odd parity, which in turn modulate the global morphology of the magnetization curves.

Once trained, the model facilitates rapid prediction of magnetization behavior for arbitrary one-dimensional spin chain topologies as described by Eq.~\eqref{eq1}, obviating the need for repeated Monte Carlo sampling. This enables a substantial reduction in computational overhead while maintaining physically consistent output.

The present study adopts the antiferromagnetic Ising model with $J = 1$ as a minimal yet non-trivial setting to evaluate the efficacy of graph-based neural networks in encoding and predicting magnetic response functions. Several avenues for future enhancement naturally follow. Incorporation of edge-level attributes encoding the magnitude and sign of coupling constants could improve the model's representational power. The integration of attention mechanisms may further enable the network to selectively emphasize physically salient substructures. Extension of the framework to higher-dimensional spin systems would allow the study of more complex forms of geometric frustration and phase behavior. Moreover, generalization to alternative spin models, including the Potts, Heisenberg, and quantum variants, would broaden the applicability of the approach and facilitate deeper exploration of magnetism through structure-based learning.

\end{document}